# Atoms can be divided into three categories: polar, non-polar and hydrogen atom

Pei-Lin You

Institute of Quantum Electronics, Guangdong Ocean University, zhanjiang 524025, China.

Since the time of Rutherford (1911) physicists and chemists commonly believed that with no electric field, the nucleus of an atom is at the centre of the electron cloud, atoms do not have permanent electric dipole moment (EDM), so that there is no polar atom in nature. In the fact, the idea is untested hypothesis. After ten years of intense research, our experiments showed that atoms can be divided into three categories: polar, non-polar and hydrogen atom. Alkali atoms are all polar atoms. The EDM of a Sodium, Potassium, Rubidium and Cesium atom in the ground state have been obtained as follows: $d_{Na}=[1.28\pm0.18]\times10^{-8}$ e.cm; $d_K=[1.58\pm0.23]\times10^{-8}$ e.cm; $d_{Rb}=[1.70\pm0.24]\times10^{-8}$ e.cm; $d_{Cs}=[1.86\pm0.27]\times10^{-8}$ e.cm. All kind of atoms are non-polar atoms except for alkali and hydrogen atoms. Hydrogen atom is quite distinct from the others. The ground state in hydrogen is non-polar atom(d=0) but the excited state is polar atom, for example, the first excited state has a large EDM: $d_H=3ea_o=1.59\times10^{-8}$ e.cm ($a_o$ is Bohr radius).

**1. Introduction**   The realization that there is one small nucleus, which contains the entire positive charge and almost the entire mass of the atom, is due to the investigations of Rutherford, who utilized the scattering of alpha particles by matter. He found that when swiftly-moving alpha particles are allowed to collide with gold atoms, they are sometimes deflected through $180^o$, implying that a strong force is at work. This experimental phenomenon led Rutherford to propose the nuclear model of the atom in 1911. The force is just the electrostatic repulsion experienced by an alpha particle when it chances to approach close to the nucleus of a gold atom. Since the time of Rutherford physicists and chemists commonly believed that in the absence of an external field, the nucleus of an atom is at the centre of the electron cloud, so that any kind of atoms does not have permanent electric dipole moment (EDM)[1]. Therefore, there is no polar atom in nature except for polar molecules.

In the fact, the scattering experiment of alpha particles by matter only showed that there is one small nucleus, its diameter is only about one ten thousandth of the atomic diameter, and the positive charge on the nucleus is precisely cancelled out by the negative charge of the electron cloud. But the idea，the nucleus of an atom is at the centre of the electron cloud (based on the spherical symmetry of the nuclear model of the atom), is untested hypothesis. R.P. Feynman once stated that "The principle of science, the definition, almost, is the following: *The test of all knowledge is experiment.* Experiment is the sole judge of scientific 'truth' ". [1]

We hope thoroughly to test the idea by precise measurement.

**2. How can we separate Polar and Non-polar atoms?**   Molecules can be divided into two categories: polar molecule (such as $H_2O$, HCl, etc) and non-polar molecule (such as $H_2$, $O_2$, etc). The electric susceptibility is [2]

$$x_e = C/C_o - 1 \qquad (1)$$

where $C_o$ is the vacuum capacitance and C is the capacitance of the capacitor filled with the material. The electric susceptibility of atoms or molecules can arise in two ways: the applied field distorts the charge distributions and so produces an induced dipole moment in each atom or molecule; the applied field tends to line up the initially randomly oriented permanent dipole moment of the polar substances. Note that the electric susceptibility caused by the orientation of polar molecules is inversely proportional to the absolute temperature:

$$x_e = N d_o^2 / 3kT \varepsilon_o = B/T \qquad (2)$$

where $d_o$ is the EDM of a molecule, k is Boltzmann constant, $\varepsilon_o$ is the permittivity of free space, N is the number density of molecules and the slope $B=Nd_o^2/3k\varepsilon_o$ is constant when N keeps a fixed density [2]. From (2) we can work out the EDM of a polar molecule

$$d_o = (3k\varepsilon_o B /N)^{1/2} \qquad (3)$$

The induced electric susceptibility due to the distortion of electronic motion in atoms or molecules is temperature independent, so the electric susceptibility is[3]

for non-polar substances $\qquad x_e = A \qquad (4)$

where A is constant. For polar substances both types of polarization, induced and orientation, are present, and the electric susceptibility is[3]

for polar substances $\qquad x_e = A + B/T \qquad (5)$

In Classical Electrodynamics by J.D.Jackson, the electric susceptibility is plotted against 1/T for polar and non-polar substances, the plot is a horizontal line for non-polar substance and an oblique line for polar substance respectively (see Fig.1)[3]. **This difference in temperature dependence offers a means of separating the polar and non-polar atoms experimentally.**

R.P. Feynman checked Eq.(5) with the orientation polarization experiment of water vapor. He plotted the oblique line from four experimental points .The table 1 gives the experimental data [1].

**Table 1**     The electric susceptibility $x_e$ of Water vapor at different temperature T

| t (℃) | T (K) | 1/T(K$^{-1}$) | $x_e$ | P(cmHg) | N(m$^{-3}$) |
|---|---|---|---|---|---|
| 120 | 393.15 | $2.5436 \times 10^{-3}$ | $400.2 \times 10^{-5}$ | 56.49 | $1.388 \times 10^{25}$ |
| 150 | 423.15 | $2.3632 \times 10^{-3}$ | $371.7 \times 10^{-5}$ | 60.93 | $1.391 \times 10^{25}$ |
| 180 | 453.15 | $2.2068 \times 10^{-3}$ | $348.8 \times 10^{-5}$ | 65.34 | $1.393 \times 10^{25}$ |
| 210 | 483.15 | $2.0698 \times 10^{-3}$ | $328.7 \times 10^{-5}$ | 69.75 | $1.395 \times 10^{25}$ |

From the ideal gas law, the average density N = P/ k T = $1.392 \times 10^{25}$ m$^{-3}$. From least-square method we obtain A=$1.8 \times 10^{-4} \approx 0$, B=1.50K, and the EDM of a H$_2$O molecule $d_{H2O}$ = $(3k\varepsilon_o B /N)^{1/2}$ =$6.28 \times 10^{-30}$ C.m = $0.393 \times 10^{-8}$e. cm, it is conform to the observed value $d_{H2O}$= $6.20 \times 10^{-30}$ C. m=$0.388 \times 10^{-8}$e. cm [2].

**If alkali atom is the polar atom and has a large EDM, a temperature dependence of the form $x_e$ =A+B/T should be expected in measuring the capacitance.** In the following two sections we shall use two different methods to find the EDM of the polar atom (for instance, Na atoms). We shall see the difference and relation between the polar atoms and polar molecules (for instance, H$_2$O molecules).

**3. Experimental method and result**    The first experiment: measuring the capacitance of Na vapor under the condition of Na saturated vapor pressure. The experimental apparatus is a closed glass container. It resembles a Dewar flask in shape. The external and internal diameters of the container are $D_1$=78.5mm and $D_2$=54.5mm. The external and internal surfaces of the container are plated with silver, respectively shown by **a** and **b** in Fig.2. These two silver layers constitute the cylindrical capacitor. The length of the two silver layers is L=26.5cm. The thickness of the glass wall is △=1.5mm. The width of the gap that will be filled by Na vapor is $H_1$=9.0mm. The magnitude of capacitance is measured by a digital capacitance meter. The precision of the meter was 0.1pF, the accuracy was 0.5% and the surveying voltage was V=1.20 volt. This capacitor is connected in series by two capacitors. One is called C' and contains the Na vapor of thickness $H_1$, another one is called C'' and contains the glass medium of thickness 2△. The total capacitance C is

$$C = C'C''/ ( C'+C'') \quad \text{or} \quad C'= C'' C / (C'' - C ) \tag{6}$$

where C'' and C can be directly measured. The experiment to measure C'' is easy. We can made a cylindrical capacitor of glass with thickness 2△ and put it in a temperature-control stove. By measuring capacitances at different temperatures, we can find C'' correspond to different temperatures . When the container is empty, it is pumped to a vacuum pressure P≤$10^{-8}$ Pa for 20 hours. The aim of the operation is to remove impurities such as oxygen adsorbed on the inner walls of the container. We measured the total capacitance C = 52.4pF and C'' = 1632pF, the vacuum capacitance is C'$_0$ =54.1 pF according to (6). The mass of a sodium sample, with purity 0.9995, is 5g and supply by Strem chemicals company USA. The next step, the Na sample is put in the container. The container is again pumped to vacuum pressure P ≤$10^{-8}$ Pa at room temperature, then it is sealed. We obtain the experimental apparatus as shown in Fig.2, and it is a glass Dewar flask filled with Na vapor and surplus liquid sample. We put the capacitor into the temperature-control stove, raise the temperature of the stove very slow and keep the temperature at $T_1$ =591.15K(318℃) for 4 hours.

When isothermal time t≥180 minutes at $T_1$ =591.15K, both the total capacitance and glass capacitance remain constants: $C_t$= 1468 pF and C'' = 8160 pF. And the capacitance of Na vapor is C'$_t$=1790 pF according to (6). It means that the readings of capacitance are obtained under the condition of Na saturated vapor pressure. So the electric susceptibility of Na vapor remains constant $x_e$ = C'$_t$/ C'$_0$ - 1=32.1. The formula of saturated vapor pressure of Na atoms is P= $10^{7.553-5395.4/T}$ psi, where 1 psi =6894.8Pa[4]. The effective range of the formula is 453K ≤T ≤1156K. Using the formula we obtain the saturated vapor pressure of Na vapor $P_{Na}$=183.9Pa at 591.15K. From the ideal gas law, the density of Na vapor $N_1$ = $P_{Na}$ /k$T_1$ =$2.253 \times 10^{22}$ m$^{-3}$.

It is well known that the electric susceptibility is of the order of $10^{-3}$ for any kind of gases, for example 0.0046 for HCl gas, 0.007 for water vapor [3]. Please notice that $x_e$ = 32.1>>1 for Na vapor and the digital meter applied the external field only with E=V/ $H_1$=$1.33 \times 10^2$V/m, and it is very weak. Our experimental result exceeded all expert's expectation!

The second experiment: investigation of the relationship between $x_e$ of Na vapor and T at a fixed density. The apparatus was the same as the preceding one but the Na vapor was at a fixed density $N_2$. In order to control the quantity of Na vapor, the container is connected to another small container that contains Na material by a glass tube from the top. These two containers are slowly heated to $T_2$=473.15K(200℃) in the stove for 3 hours and the designed experimental container is sealed. The capacitance C of the capacitor was still measured by the digital meter and the vacuum capacitance $C_{20}$=44.0pF. Its length is $L_2$=23.0 cm and the plate separation is $H_2$=8.5mm. From the formula of saturated vapor pressure we obtain $P_{Na}$=0.9736 Pa at $T_2$=473.15K. Similarly, the density of Na atoms $N_2 = P_{Na}/kT_2 = 1.490 \times 10^{20}$ m$^{-3}$. By measuring the electric susceptibility of Na vapor at different temperature, we obtain $x_e$=A+B/T≈B/T, where the intercept A≈0 and the slope of the line B=126.6 ±3.8 (K). The experimental results are shown in Fig.3. Table 2 gives a complete experimental data.

**Table 2 The electric susceptibility $x_e$ of Na vapor at different temperature T**

| t (℃) | 114 | 122 | 130 | 138 | 146 | 154 | 164 | 174 | 186 |
|---|---|---|---|---|---|---|---|---|---|
| T(K) | 387.15 | 395.15 | 403.15 | 411.15 | 419.15 | 427.15 | 437.15 | 447.15 | 459.15 |
| 1/T(×10$^{-3}$) | 2.5830 | 2.5307 | 2.4805 | 2.4322 | 2.3858 | 2.3411 | 2.2875 | 2.2364 | 2.1779 |
| C(pF) | 58.8 | 58.5 | 58.3 | 58.0 | 57.7 | 57.4 | 57.1 | 56.9 | 56.6 |
| $x_e$ | 0.33636 | 0.32955 | 0.3250 | 0.31818 | 0.31136 | 0.30455 | 0.29773 | 0.29318 | 0.28636 |

where $C_0$= 44.0pF, $N_1$=1.490×10$^{20}$ m$^{-3}$. From least-square method we obtain B= 126.6K and A=0.0096≈0, where the surveying voltage V=1.20 volt. The digital meter applied the external field only with E=V/$H_2$=1.4×10$^2$V/m, and it is very weak. From Eq. (11) we work out $d_{Na}$ = 2.048×10$^{-29}$C.m= 1.280×10$^{-8}$e.cm＞$d_{H2O}$!

The third experiment: measuring the capacitance of Na vapor at various voltages (V) under the fixed density $N_2$ and a fixed temperature $T_3$. The apparatus was the same as the preceding one ($C_{30}$= $C_{20}$=44.0 pF) and keep at $T_3$=303.15K (30℃). When V=$V_1$≤0.4volt, the capacitance of Na vapor $C_1$=216pF or $x_e$ =3.91, it is constant. With the increase of voltage, the capacitance decreases gradually. When V=$V_2$= 400 volt, the capacitance of Na vapor $C_2$=47.0pF or $x_e$= 0.0682, it approaches saturation. The measured result showed that the saturation polarization of the Na vapor is obvious when E≥$V_2$/$H_2$=4.7×10$^4$V/m [18].

**4. Theory and interpretation** We obtain the first formula of atomic EDM( see discussion ④)

$$d_{atom} = (C - C_o)V / L(a)SN \qquad (7)$$

where L(a) is the Langevin function. We work out L(a)=0.082 in the first experiment. L(a)=0.082 shows that only 8.2% of Na atoms are lined up with the direction of the field in the first experiment. Substituting the values: $S_1$= 5.50×10$^{-2}$ m$^2$, $N_1$=2.253×10$^{22}$ m$^{-3}$, V=1.20volt and C - $C_o$=$C'_t$ - $C'_0$ = 1735.9pF, we work out

$$d_{Na} = (C - C_o)V / L(a) S_1 N_1 = 2.050 \times 10^{-29} \text{C.m} = 1.281 \times 10^{-8} \text{e.cm} \qquad (8)$$

The statistical error of the measured value is △$d_1$/d＜0.12, considering all sources of systematic error: △$d_2$/d＜0.08, and the combination error is △d/d＜0.15. We find that

$$d_{Na} = [2.05 \pm 0.30] \times 10^{-29} \text{C.m} = [1.28 \pm 0.18] \times 10^{-8} \text{e.cm} \qquad (9)$$

From d= ( 3k $\varepsilon_o$B / $N_2$ )$^{1/2}$, note that k= dE/a$T_1$= dV/a$T_1H_1$, we obtain **the second formula of atomic EDM**

$$\mathbf{d_{atom}= ( 3k \varepsilon_o B / N_2 )^{1/2}= 3 V \varepsilon_o B/ aT_1H_1N_2} \qquad (10)$$

Substituting the values: $T_1$= 591.15k, $H_1$=9.0mm, V=1.20volt, B=126.6K, a=0.2485( see discussion ④) and $N_2$ =1.490×10$^{20}$ m$^{-3}$, from Eq. (10) we work out

$$d_{Na} = \mathbf{3 V \varepsilon_o B/ aT_1H_1N_2} = 2.048 \times 10^{-29} \text{C.m} = 1.280 \times 10^{-8} \text{e.cm} \qquad (11)$$

**Using two different methods and different experimental data we obtain the same result, it proved that the data are reliable, and the EDM of an Na atom has been measured accurately. Although above calculation is simple, but no physicist completed the calculation up to now!**

**5. Discussion**

①The shift in the energy levels of an atom in an electric field is known as the Stark effect. Normally the effect is quadratic in the field strength, but the first excited state in hydrogen exhibits an effect that is linear in the strength. This result shows that the hydrogen atom (the quantum number n=2 ) has very large EDM, $d_H$=3e$a_o$=1.59×10$^{-8}$e.cm ($a_o$ is Bohr radius)[5]. L.D. Landay once stated that "The presence of the linear effect means that, in the unperturbed state, the hydrogen atom has a dipole moment"[6]. The alkali atoms having only one valence electron in the outermost shell can be described as hydrogen-like atoms[5]. The quantum number of the ground state alkali atoms are n≥2 rather than n=1( this is 2 for Li, 3 for Na, 4 for K, 5 for Rb and 6 for Cs), as the excited state in hydrogen. In 2000 I conjecture that the ground state neutral alkali atoms may have large EDM of the order of e $a_o$ [9].Due to the EDM of an atom is extremely small and we applied several ingenious experimental techniques [10].

②The formula $d_{atom}$ =(C－$C_o$)V/L(a)SN can be justified easily. The dipole moment of an Na atom is d = e r. N is the number of Na atoms per unit volume. L(a) is the percentage of Na atoms lined up with the field in the total number. When an electric field is applied, the Na atoms tend to orient in the direction of the field as dipoles. On the one hand, the change of the charge of the capacitor is △Q=(C－$C_0$)V. On the other hand, due to the volume of the capacitor is SH, the total number of Na atoms lined up with the field is SHNL(a).The number of layers of Na atoms which lined up with the filed is H/r. Because inside the Na vapor the positive and negative charges cancel out each other, the polarization only gives rise to a net positive charge on one side of the capacitor and a net negative charge on the opposite side. Then the change of the charge is△Q=SHN L(a)e / (H/r) = SN L(a)d = (C－$C_0$)V, so the EDM of an Na atom is d = (C－$C_0$)V/ SN L(a).

③From 1ev=kT we get T=11594K, in the temperature range of the experiment 303K≤T≤591K, kT<<1ev, so the capacitance change ( $C'_t$－$C'_0$ ) entirely comes from the contribution of Na atoms in the ground state.

④The local field acting on a molecule in a gas is almost the same as the external field **E**[2]. The electric susceptibility of a gaseous polar dielectric is[7]

$$x_e = NG + N d_o L(a)/ \varepsilon_o E \qquad (12)$$

where a = $d_o$ E /kT, $d_o$ is EDM of a molecule, G is the molecule polarizability. The Langevin function is

$$L(a) = [(e^a + e^{-a})/(e^a - e^{-a})] - 1/a \qquad (13)$$

The Langevin function L(a) is equal to the mean value of cos θ ( θ is the angle between **$d_o$** and **E**) [7]:

$$<\cos \theta> = \mu \int_0^\pi \cos \theta \exp(d_o E \cos \theta /kT) \sin \theta \, d\theta = L(a), \quad \mu = [\int_0^\pi \exp(d_o E \cos \theta /kT) \sin \theta \, d\theta]^{-1} \qquad (14)$$

where μ is a normalized constant. This result shows that L(a) is the percentage of polar molecule lined up with the field in the total number. When a<<1 and L(a)≈a/3, when a>>1 and L(a)≈1[7].

The next step, we will consider how this equation is applied to Na atoms. Due to the atomic polarizability of Na atoms is G= 24.1×$10^{-30}$ $m^3$[8], the number density of Na atoms N≤$10^{23}$ $m^{-3}$ and the induced susceptibility A=NG≤2.41×$10^{-6}$ can be neglected. In addition, the induced dipole moment of Na atoms is $d_{int}$ =G $\varepsilon_o$E[8], due to E≤$10^5$v/m in the experiment, then $d_{int}$≤2.13×$10^{-35}$ C.m can be neglected, and we obtain

$$x_e = Nd L(a)/ \varepsilon_o E \qquad (15)$$

where d is the EDM of an Na atom and N is the number density of Na vapor. L(a)= < cos θ > is the percentage of Na atoms lined up with the field in the total number. Note that E=V/H and $\varepsilon_o$= $C_o$ H / S, leading to

$$C - C_o = \beta L(a)/a \quad \text{or} \quad x_e = \beta L(a)/a \, C_o \qquad (16)$$

where β = S N $d^2$/kTH is a constant. **This is the polarization equation of Na atoms**. Due to a=d E/kT= dV/kTH we obtain **the first formula of atomic EDM**

$$\mathbf{d_{atom} =(C－C_o )V / L(a)SN} \qquad (7)$$

In order to work out **L(a)** and **a** of the first experiment, note that in the third experiment when the field is weak ($V_1$=0.4V), $a_1$ <<1 and L($a_1$)≈$a_1$ /3. From Eq.(16): $C_1$－$C_{30}$=216－44= β /3 and β =516pF. When the field is strong ($V_2$=400V), $a_2$ >>1 and L($a_2$)≈1, $C_2$－$C_{30}$= L($a_2$) β /$a_2$. We work out $a_2$ = β L($a_2$)/($C_2$－$C_{30}$)= 171>>1, L($a_2$)=0.9942. Due to **a**=dV/kTH, so **a**/$a_2$=V$T_2$$H_2$/$T_1$$H_1$$V_2$, Substituting the corresponding values, we obtain **a**= 0.2485 and L(a)=0.082. Notice that we deduced Eq(16) from the formula of the parallel-plate capacitor $\varepsilon_o$ = $C_o$H /S, so the cylindrical capacitor must be regarded as a equivalent parallel-plate capacitor with the plate area S= $C_o$ H/ $\varepsilon_o$. In the first experiment the equivalent plate area $S_1$= $C'_0$ $H_1$/ $\varepsilon_o$ =5.50×$10^{-2}$ $m^2$.

⑤Accurate measurements of the EDM of an Cesium (Cs) , Rubidium (Rb) and Potassium(K) atom in ground state have been carried out. Similar results have been obtained as follows[11-21]:

$d_K$ = [1.58 ±0.23]×$10^{-8}$e. cm ( see Fig.4, arXiv: 0908.3955V2)

$d_{Rb}$= [1.70 ±0.24]×$10^{-8}$e. cm ( see Fig.5, arXiv: 0810.0770 )

$d_{Cs}$= [1.86 ±0.27]×$10^{-8}$e. cm ( see Fig.6, arXiv: 0809. 4767 )

The capacitance of the Hg vapor at different temperature was measured and we obtain B≈0.0K. Hg atoms are non-polar atoms. The experimental results are shown in Fig.6.

**Acnowledgement**

The author thank to Prof. Song-Hao Liu, Dang Mo, Xiang-You Huang and Wei-Min Du, thank to Yu-Sheng Zhang, Rui-Hua Zhou, Zhao Tang , Xue-Ming Yi, Yi-Quan Zhan, Jia You and Xin Huang for their help in the work.

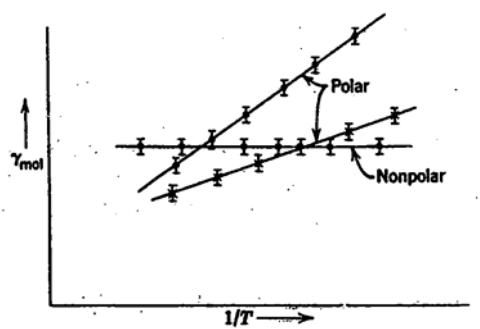

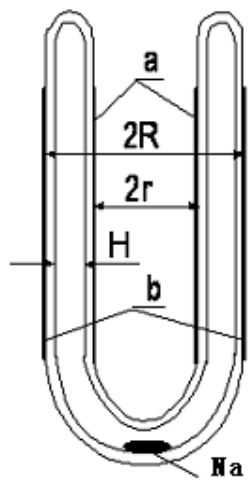

**Fig.1** It is Fig. 4.10 in Classical Electrodynamics by J.D.Jackson. Variation of the susceptibility with temperature: the plot is a horizontal line for non-polar substance, whereas an oblique line for polar.

**Fig.2** The apparatus is a glass Dewar flask filled with Na vapor and surplus liquid sample. Silver layers **a** and **b** build up the cylindrical glass capacitor (not to scale).

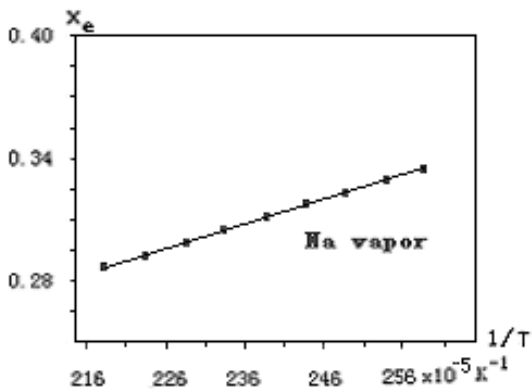

**Fig.3** The curve showed that the electric susceptibility $x_e$ of Na vapor varies inversely proportional to the absolute temperature T at a fixed density. $x_e = A+B/T$, where the slope of the line is B=126.6K, and A=0.0096≈0.

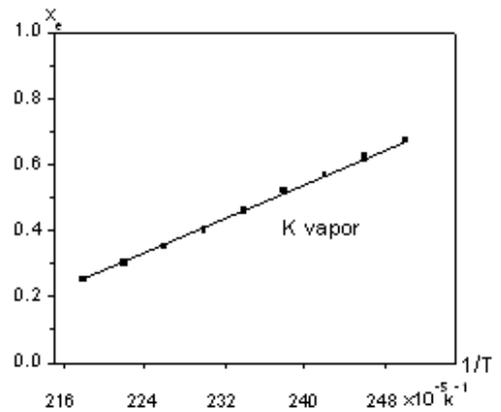

**Fig.4** The curve showed that the electric susceptibility $x_e$ of K vapor varies inversely proportional to the absolute temperature T at a fixed density. $x_e = A+B/T$, where A≈0 and the slope of the line is B=188K.

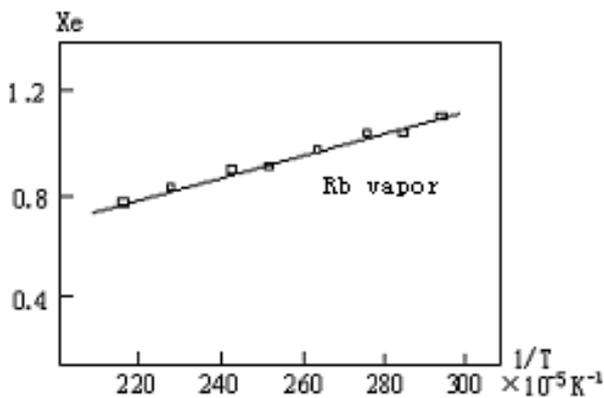

**Fig.5** The curve showed that the electric susceptibility $x_e$ of Rb vapor varies inversely proportional to the absolute temperature T at a fixed density, and $x_e = A+B/T \approx B/T$, where the slope is B=380K.

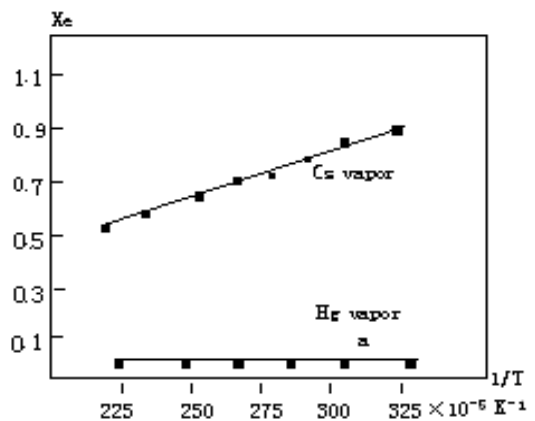

**Fig.6** The temperature dependence of the susceptibility of Cs or Hg vapor. For Cs vapor the slope B=320K, while for Hg vapor B≈0.0K. Note $d_{Cs} > d_{Rb}$, but the density $N_{Cs} \ll N_{Rb}$, so Cs vapor's slope is less than Rb vapor's.

6